\documentstyle[aps,epsf]{revtex}
\input epsf.sty
\begin{document}
\draft
\flushbottom
\twocolumn[
\hsize\textwidth\columnwidth\hsize\csname @twocolumnfalse\endcsname
\title{Non-Fermi liquid behavior and Griffiths phase in {\it f}-electron
compounds}
\author{A.~H.~Castro~Neto $^1$, G.~Castilla $^1$ and B.~A.~Jones $^2$}
\address{$^1$ Department of Physics,
University of California,
Riverside, CA 92521 \\
$^2$ IBM Almaden Research Center, San Jose, CA 95120-6099 }
\date{\today}
\maketitle
\tightenlines
\widetext
\advance\leftskip by 57pt
\advance\rightskip by 57pt

\begin{abstract}
We study the interplay among disorder, RKKY
and Kondo interactions in {\it f}-electron alloys. We argue that
the non-Fermi liquid behavior observed in these systems
is due to the existence of a Griffiths
phase close to a quantum critical point. The existence of
this phase provides a unified picture of a large class of materials.
We also propose new experiments that can test these ideas.
\end{abstract}
]
\begin{figure}
\epsfysize5cm
\hspace{1cm}
\epsfbox{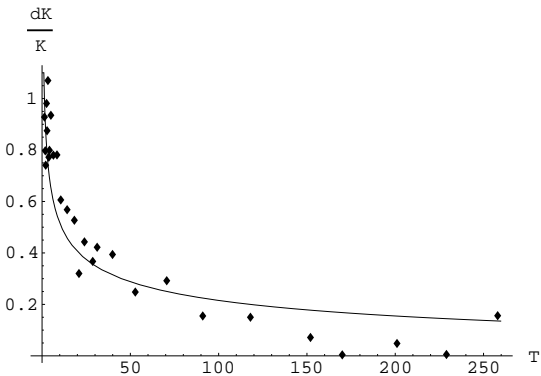}
\caption{Mean square deviation of the Knight shift as a function of
temperature given in (\ref{finalresult}).
Diamonds: experiments for UCu$_4$Pd from ref.[7].}
\label{doug}
\end{figure}

\narrowtext
\tightenlines
The observation of non-Fermi-liquid (NFL) exponents in the thermodynamic
and transport properties of {\it f}-electron alloys has stimulated
considerable
interest in the study of these materials \cite{staba}. The alloys in which
NFL behavior is observed fall into two categories:
1) Kondo hole systems, in which the {\it f}-electron atoms (R) are replaced
by non-magnetic metallic atoms (M) according to the formula
R$_{1-x}$M$_{x}$, and 2) disordered ligand systems, in which the metallic
atoms are substituted for a different metallic atom according
to the formula R(M1)$_{1-y}$(M2)$_y$. Notice that due to
alloying these compounds have a high probability of being disordered.
That disorder is indeed a very important factor in bringing
about the
NFL behavior in these compounds has been shown in
recent experiments\cite{thompson}. This is
in addition to the fact that  most of these systems are
close to a phase transition.
Then, we claim that the NFL
properties of these compounds
are a consequence of the competition between the
intra-site Kondo and the
inter-site Ruderman-Kittel-Kasuya-Yosida (RKKY) interactions
taking place
in the midst of a disordered environment.
If disorder were not present, there are two possibilities:
the compound will have long-range
magnetic order when the RKKY interaction is sufficiently large
compared with the Kondo interaction, or the compound will
be paramagnetic
due to the quenching of the magnetic moments of
the rare earth atoms.
However, the experimental observations show that
the  NFL behavior generally appears between these two phases \cite{staba}.
Several proposals
have been put forward as the possible explanations for
the NFL behavior. A possible scenario is based on
single impurity models with particular symmetries
such as
multichannel Kondo effect of
magnetic \cite{nozieres} and electric origin \cite{cox,coxagain}.
Another possible scenario attributes the NFL behavior to
proximity to a quantum critical
point\cite{hertz,georges}. Recently, a route to NFL behavior
that emphasizes
a disorder-driven  mechanism, known as ``Kondo disorder'',
has been
suggested \cite{bernal,dobro}.
All of these proposals have had partial success
in explaining some of the experiments\cite{staba}. In particular,
conformal invariance scaling gives a good description
of the dynamic susceptibility $\chi ''(\omega, T)$
and the electrical resistivity in UCu$_{5-x}$ Pd$_{x}$
\cite{MAronson};
a Kondo disorder model can explain the temperature dependence of
the electrical resistivity\cite{dobro};
and the phase diagram
of spin glass models is
qualitatively similar to the one observed in these alloys\cite{georges}.
Nevertheless,
there is no final word as to the origin of the
NFL behavior in these alloys.
Here, we propose a framework that incorporates
what we believe are
the essential aspects of the problem: disorder and
the competition
between RKKY and Kondo effect. In this framework the
presence of disorder
leads to the coexistence of a metallic
(paramagnetic) phase with a {\it granular} magnetic phase.
We show that this coexistence phase is equivalent to the
{\it Griffiths phase} of dilute magnetic systems \cite{grif}.
In our scenario, we have two electronic fluids: one of them
is quenched by the Kondo interaction,
behaving as a Fermi liquid; and
the other is dominated by the RKKY interaction,
leading to ordered regions of two level systems.
We have, therefore,
an inhomogeneous environment which is brought about
by disorder. This scenario is reminiscent of the one
found in compensated doped semiconductors
(Si:P, Ge:Sb)\cite{Fisher,lang,bhatt}.
In these
systems, disorder leads to local density fluctuations which
result in the formation of magnetic
moments.
The Griffiths phase is characterized by the formation of
rare strongly-coupled magnetic clusters
which have large susceptibilities.
In this phase the thermodynamic functions show essential singularities
with strong effects at low temperatures.
At these low temperatures, clusters of interacting magnetic moments
can be thought of
as ``giant" spins which can tunnel over classically forbidden
regions.
In the Griffiths phase magnetic clusters with $N$ spins have a relaxation
time
which is given by \cite{literature} (we use units such that $\hbar=k_B=1$)
\begin{eqnarray}
\tau_R = \omega_0^{-1} e^{N \zeta} \, ,
\label{taur}
\end{eqnarray}
where $\omega_0$ is an attempt frequency and $\zeta$ is a
characteristic parameter that we discuss below.
Due to cluster formation in the paramagnetic
phase of these systems, we have the following predictions for
the thermodynamic functions:
\begin{eqnarray}
\gamma \equiv
C_V/T &\propto& \left[\chi(T)\right]_{av} \propto T^{-1+\lambda} \; ,
\nonumber
\\
\left[\chi_{nl}(T)\right]_{av} &\propto& T^{-3+\lambda} \; ,
\nonumber
\\
\left[\chi_L(\omega)^{"}\right]_{av} &\propto& \omega^{-1+\lambda}
\tanh(\omega/T)  \; ,
\nonumber
\\
T^{-1}_1(\omega) &\propto& \omega^{-2+\lambda} T \tanh(\omega/T) \; ,
\nonumber
\\
\delta \chi(T)/\chi(T) &\propto& T^{-\lambda/2} \; ,
\label{finalresult}
\end{eqnarray}
where $[...]_{av}$ means average over disorder.
$C_V$ is the specific heat; $\chi(T)$ is the static susceptibility;
$\chi_{nl}(T)$ is the non-linear static susceptibility;
 $\chi_L(\omega)$ is
the local frequency dependent susceptibility;
$1/T_1$ is the NMR relaxation rate; and $\delta \chi(T)$ is the
mean square deviation of the susceptibility due to the distribution
of susceptibilities in the system \cite{huse}. Here
$\lambda=-\zeta^{-1} \ln(c)$, where $c$
 denotes the density of the spins.
The Griffiths phase is characterized by $\lambda <1$
so that the susceptibilities diverge at zero temperature.
We propose that Griffiths singularities dominate the
physics of the system at low temperatures
 leading to NFL behavior.
Let us note here that power-law behaviors
for $\gamma$ and $\chi$ have also been
obtained by other researchers
using different approaches\cite{MAronson,lang,bhatt}.
Notice that the logarithmic behavior observed
in some NFL compounds \cite{staba} can as well be
fitted by small power laws ($\lambda \approx 1$).
Also, it follows from (\ref{finalresult}) that
NFL systems should have {\it positively} divergent
non-linear susceptibilities ($\lambda<3$).
Indeed, U$_{0.9}$Th$_{0.1}$Be$_{13}$ shows a tendency to a
positively divergent susceptibility \cite{aliev},
in contrast to the usual negative divergence
of the paramagnetic susceptibility of UBe$_{13}$.
Systems like UCu$_{5-y}$Pd$_y$ show even stronger divergent
behavior and can be considered ``deep" inside of the Griffiths
phase. Recent neutron scattering experiments show that the
imaginary part of the susceptibility, the specific heat and the
static susceptibility can be exactly fitted by the result
(\ref{finalresult})
with $\lambda = 2/3$ \cite{aronson}.
To compare this result with NMR and
$\mu$SR,
 we calculated the variation of the Knight shift, $\delta K/K \propto
\delta \chi/\chi$,
for the same material \cite{bernal}. Our result is shown in Fig.\ref{doug}
for
$\lambda =2/3$.
The agreement between theory and experiment is very good.
Our predictions are robust in the sense that {\it all} the results in
(\ref{finalresult}) have to be self-consistent.
Moreover, we have a definite prediction for the NMR relaxation rate
$1/T_1(\omega)$, which should be largely frequency dependent.
We also predict that under pressure the exponent
$\lambda$ should change inside of the Griffiths phase \cite{bob}.
It would be interesting to plot the logarithmic derivative
of the susceptibility (and/or specific heat) as a function of
temperature and pressure, $\lambda(T,P) = 1 + \partial \ln
\chi(T,P)/\partial
\ln T$, for various NFL systems to verify our predictions.
The main characteristics
of {\it f}-electron systems studied here are: interplay between RKKY and
Kondo
effects; magnetic anisotropy; and disorder due to alloying.
It is well known that {\it f}-electrons systems are
characterized by their strong magnetic properties. This magnetism
arises from the crystal field interactions and the
strong spin-orbit coupling;
in some of these materials, the large {\it f}- electronic
clouds lead to an anisotropy comparable in magnitude to
the exchange energy. In addition to this ion anisotropy, one
expects a Dzyaloshinsky-Moriya (DM) exchange interaction \cite{dzmo}
generated by the spin-orbit coupling.
This type of interaction is only forbidden in highly symmetric
situations which are not commonly realized in alloys with very
complex unit cells \cite{yosida}. That this is the case, it
can be seen from
the coexistence of weak parasitic ferromagnetism within the
antiferromagnetic phase in some materials such as
R-Cr0$_3$ systems\cite{koeler}.
We believe that this DM interaction accounts for the
recent neutron scattering data in
the NFL compound CeCu$_{6-y}$Au$_y$ \cite{vonnew}. Consequently
the RKKY and Kondo interactions will be strongly anisotropic
in the {\it f}-electron systems. The simplest Hamiltonian
that describes the situation above is the anisotropic Kondo model
\begin{eqnarray}
H = H_e + \sum_{n} \left[J^z_{n} S^z_n s^z_n + \frac{J^{\perp}_n}{2}
\left(S^+_n s^-_n + S^-_n s^+_n\right)\right]
\label{kondolat}
\end{eqnarray}
where the sum is over all the R atoms with spin ${\bf S}_n$.
$H_e$ is the conduction electron Hamiltonian that can be obtained from
band structure calculations, and
$s^a_n = \sum_{\sigma,\sigma'} c^{\dag}_{n,\sigma} \tau^a_{\sigma,\sigma'}
c_{n,\sigma'}$ with $a=x,y,z$ is the conduction electron spin
($\tau^a$ are Pauli matrices). The main difference between Kondo
hole and disordered ligand systems is the alloying process. In the Kondo
hole
the magnetic atom is replaced which leads to a reduction of the
number of magnetic moments. Moreover, the
substitute atom has a different size from the
original R atom which will lead to local changes in the lattice
structure. In disordered ligand systems the lattice spacing is also
affected by the substitution of the ligand atoms.
As is well known, hybridization matrix elements between
the conduction band and the {\it f}-electron system are exponentially
sensitive to changes such as the type of lattice and characteristics
of the substitute atom, and this will in turn affect the local values
of the exchange constants in (\ref{kondolat}). Therefore, alloying
leads to a situation where local variations result in parts of the
sample having larger exchange interactions than others.
The magnetism in these systems can be understood
by looking at the effective interaction
between the R atoms due to the conduction electrons. The magnetic
Hamiltonian in second order perturbation theory relative to the
free electron problem is
\begin{eqnarray}
H_M \approx \sum_{n,m,a,b} S^a_n \, \Gamma^{a,b}_{n,m} \, S^b_m
\label{RKKY}
\end{eqnarray}
where $\Gamma^{a,b}_{n,m}$ is the RKKY interaction mediated by
the conduction electrons.
In the limit of large anisotropy,
$J^z >> J^{\perp}$, we have $\Gamma^{z,z} \sim {\cal O}[(J^z)^2/E_F]$
where $E_F$ is the Fermi energy.
In a clean $d$-dimensional system this interaction decays like
$1/r^d$. In the presence of disorder, however, it decays
like $e^{-r/\ell}$ where $\ell$ is the spin-orbit diffusion length
\cite{elihu}.
Since we are dealing with disordered systems the effective interaction
$\Gamma^{a,b}_{n,m}$ is actually short-ranged. The important point here
is that the ordering temperature of the magnetic system, $T_c$,
is of order of the exchange interaction, that is,
$T_c(J^z) \propto \Gamma \propto (J^z)^2/E_F$.
The critical ordering temperature is not the only energy
scale in the problem since the Kondo effect
can take place and quench the magnetic moment.

Consider for instance
a particular position in the system, say $n=M$, where, due to the alloying,
$J^z_M$ is much larger than average. For simplicity we disregard
all the other sites and look at the physics at this particular
site. The Hamiltonian of interest can be written as $H_M = H_0 + H_I$,
where
\begin{eqnarray}
H_0 &=& H_e + J^z_{M} S^z_n \left(n_{M,\uparrow} - n_{M,\downarrow}\right)
\nonumber
\\
H_I &=& \frac{J^{\perp}_M}{2} \left(S^+_M s^-_M + S^-_M s^+_M\right)
\label{kondo}
\end{eqnarray}
and in the limit of large anisotropy we treat $H_I$ as a perturbation
of $H_0$. Observe that $H_0$ can be easily diagonalized in the
$S^z$ basis. If the R atom is in a state such that $S^z$
is $\Uparrow$ ($\Downarrow$) the energy of
the system is minimized
by making a bound state with an electron with spin $\downarrow$
($\uparrow$), respectively. The bound state
energy is just the Kondo temperature
\begin{eqnarray}
T_{K}(M) = W \, \, e^{-1/(N(0) J^z_M)}
\label{kondotemp}
\end{eqnarray}
where $N(0)$ is the renormalized
density of states at the Fermi energy, and $W$ is
the conduction electron bandwidth.
This situation is similar to the approach to NFL
behavior which takes into account a disordered distribution of
Kondo temperatures \cite{bernal,dobro}. The ground state of
$H_0$ is doubly degenerate corresponding to the
spin configurations $|\Uparrow,\downarrow\rangle$
and $|\Downarrow,\uparrow\rangle$.
Application of first order perturbation theory in $H_I$
shows that the singlet state $|\Uparrow,\downarrow\rangle
- |\Downarrow,\uparrow\rangle$ is lower in energy than the
triplet state $|\Uparrow,\downarrow\rangle
+ |\Downarrow,\uparrow\rangle$ by an amount $J^{\perp}_M$. In this case
$H_I$ acts as a {\it transverse field} and lifts the degeneracy of
the bound state. In fact,
using the bosonized version \cite{ek} of
Hamiltonian (\ref{kondo}) and mapping the problem into the dissipative
two-level system \cite{leggett}, we can show that tracing out the
electrons of the problem results in a
$S^x_M$ operator for the R atoms spin degrees
of freedom \cite{saleur}.
The competition between the RKKY interaction and the
Kondo effect can be understood in
terms of the two relevant energy scales, that is, $T_{K}(M,J^z)$ and
$T_c(J_z)$:
{\it i}) if $T_{K}(M)>>T_c$ as we lower the temperature of
the system below $T_{K}(M)$ the local magnetic moment is quenched
and order is inhibited; {\it ii}) if $T_{K}(M)<<T_c$
and the temperature is lowered below $T_c$ there is local
magnetic order and the Kondo effect is suppressed\cite{jones}.
If we now take into account the magnetic moments that are
left to interact via RKKY, as in (\ref{RKKY}), together with
the sites which are quenched by the Kondo effect, as in
(\ref{kondo}), we see that the magnetism of the original
problem in the limit where $J^z_n>>J^{\perp}_n$ is
described by
\begin{eqnarray}
H_{eff} \approx \sum_{\langle i,j \rangle} \Gamma_{i,j}
S^z_i S^z_j + \sum_i t_i S^x_i
\label{transising}
\end{eqnarray}
where the brackets $\langle i,j \rangle$ imply nearest neighbor
coupling. Notice that $\Gamma_{i,j} \sim {\cal O}((J^z)^2/E_F)$ and
$t_i \sim {\cal O}(J^{\perp})$ are random (but inter-correlated)
variables dependent on the alloying and
lattice structure. Thus
we have mapped our problem into the random Ising model in
a random transverse magnetic field \cite{literature}.
The phase diagram of this model follows: at small doping the
RKKY interaction dominates and the system can order magnetically
(the ordered phase can be antiferromagnetic, ferromagnetic or
spin glass \cite{georges} depending if the mean value of
$\Gamma_{i,j}$ in (\ref{transising}) is positive, negative
or null, respectively).
With increasing doping the quantum fluctuations grow
due to the Kondo effect and the bulk critical temperature
decreases until it vanishes for some critical value of doping.
At this quantum critical point the system percolates. For large
values of doping, inside of the paramagnetic phase, only finite
clusters of magnetic atoms can be found. Among these clusters
there are some rare ones which are large and strongly coupled.
Within these clusters the spins behave coherently as a ``giant"
spin or a magnetic grain. We can describe the cluster in terms
of a ``classical" degree of freedom which can be parametrized by
Euler angles $(\theta,\phi)$. The classical energy,
$E(\theta,\phi)$,
will have at least two minima
due to the original degeneracy of the magnetic ground state.
In the simplest
of cases,
$E(\theta,\phi)$
can be written in terms of a classical
spin with $X$-easy axis and $XY$-easy plane,
\begin{eqnarray}
E(\theta,\phi) = N
\left(-\epsilon_{\perp} + \epsilon_{||} \sin^2 \phi \right) \sin^2 \theta
\label{eneclass}
\end{eqnarray}
where $\epsilon_{\perp} > \epsilon_{||}>0$ are the anisotropy
energies perpendicular and parallel to the easy axis, respectively.
These energies depend on the microscopic coupling constants in
(\ref{transising}).
Observe that the energy has two minima at $(\pi/2,0)$ and
$(\pi/2,\pi)$ with an energy barrier between them. When the
temperature is higher than the barrier height the cluster
is thermally activated and behaves classically. At lower temperatures
the cluster can undergo quantum tunnelling between the two minima.
Using standard instantons methods \cite{prok},
we can calculate the parameters that appear in (\ref{taur}):
\begin{eqnarray}
\omega_0 &=& 2 \sqrt{\epsilon_{||} \epsilon_{\perp}}
\nonumber
\\
\zeta &\approx& \ln\left(\frac{1+\sqrt{\epsilon_{||}/\epsilon_{\perp}}}{
1-\sqrt{\epsilon_{||}/\epsilon_{\perp}}}\right) \, .
\label{aniso}
\end{eqnarray}
Notice that in the absence of anisotropy ($\epsilon_{||}=\epsilon_{\perp}$)
tunnelling cannot occur ($\zeta \to \infty$ and $\tau_R \to \infty$), as
expected.
Then, the low energy physics for a cluster
$\Omega$ with $N$ spins
reduces to the Hamiltonian
\begin{eqnarray}
H_{\Omega(N)} = \Delta_{\Omega(N)} \,\, \tau^x
\label{tunnel}
\end{eqnarray}
where $\Delta_{\Omega(N)} = 1/\tau_R$ is the tunnelling energy
and is given in (\ref{taur}), and it can be related to the anisotropy
energies by (\ref{aniso}). Using (\ref{tunnel}) and averaging
over cluster with different sizes, we arrive at the predictions
given in (\ref{finalresult}) \cite{huse}.
In conclusion, we propose that the NFL behavior observed in
{\it f}-electron systems can be attributed to the existence of
Griffiths singularities close to the quantum critical point.
These singularities have their origin on the interplay between
the RKKY and Kondo interactions in the presence of magnetic
anisotropy and disorder. These conclusions have similarities
with those discussed recently by Sachdev\cite{sachdev}.
We were also able to map the disordered
Kondo lattice problem into the random Ising model in a random
transverse magnetic field where disorder is correlated. In
the paramagnetic phase of this Ising model the physics of clusters
can be understood in terms of the quantum tunnelling of intrinsic
magnetic grains which are described by a classical spin model.
At low temperatures the spin degrees of freedom of a
magnetic grain can tunnel over classically
forbidden regions and
at finite temperatures they can be thermally activated.
At very low temperatures the problem reduces
to a two-level system problem which, when appropriately averaged
over disorder, leads to the Griffiths singularities and
to the predictions in (\ref{finalresult}). This
Griffiths phase will depend strongly on the type of lattice
structure and value of the local microscopic exchange constants.
This would explain why
systems like CeCu$_{6-y}$Au$_y$
have to be fine tuned for NFL behavior to be observed,
while others, like Th$_{1-x}$U$_x$Pd$_3$,
have large regions of NFL behavior. It is indeed possible,
as in 1D systems \cite{literature}, that Griffiths behavior
can extend over large regions of doping \cite{exception}.
We acknowledge D.~Maclaughlin for providing us with the experimental
data and M.~C.~de Andrade, M.~Aronson, W.~Beyermann, R.~Bhatt, D.~Cox,
V.~Dobrosavljevi\'c,  D.~Huse, M.~B.~Maple, P.~Nozi\`eres
H.~Rieger and P.~Young for illuminating discussions. A.~H.~C.~N.
acknowledges support from the A.P.~Sloan foundation
and support provided by the DOE for research at Los Alamos National
Laboratory.

{\it Note added}.-- After this paper was completed, experimental indications of
possible Griffiths phase behavior have been reported\cite{marcio}.

\end{document}